\documentclass[reprent,aps,prb,twocolumn,
superscriptaddress,nofootinbib]{revtex4-2}
\usepackage{amsmath,amssymb}
\usepackage{graphicx}
\usepackage{bm}
\usepackage{multirow}
\usepackage{braket}

\usepackage{color}

\begin{document}

\title{
Anisotropic spin model and multiple-$Q$ states in cubic systems
}
\author{Ryota Yambe}
\email{yambe@g.ecc.u-tokyo.ac.jp}
\affiliation{Department of Applied Physics, The University of Tokyo, Tokyo 113-8656, Japan }
\author{Satoru Hayami}
\email{hayami@phys.sci.hokudai.ac.jp}
\affiliation{Graduate School of Science, Hokkaido University, Sapporo 060-0810, Japan}

\begin{abstract}
Multiple-$Q$ states manifest themselves in a variety of noncollinear and noncoplanar magnetic structures depending on the magnetic interactions and lattice structures. 
In particular, cubic-lattice systems can host a plethora of multiple-$Q$ states, such as magnetic skyrmion and hedgehog lattices. 
We here classify momentum-dependent anisotropic exchange interactions in the cubic-lattice systems based on the magnetic representation analysis.
We construct an effective spin model for centrosymmetric cubic space groups, $Pm\bar{3}m$ and $Pm\bar{3}$, and noncentrosymmetric ones, $P\bar{4}3m$, $P432$, and $P23$: The former include the symmetric anisotropic exchange interaction, while the latter additionally include the Dzyaloshinskii-Moriya interaction.  
We demonstrate that the anisotropic exchange interaction becomes the origin of the multiple-$Q$ states by applying the anisotropic spin model to the case under $Pm\bar{3}$. 
We show several multiple-$Q$ instabilities in the ground state by performing simulated annealing.
Our results will be a reference for not only exploring unknown multiple-$Q$ states but also understanding the origin of the multiple-$Q$ states observed in both noncentrosymmetric and centrosymmetric magnets like EuPtSi and SrFeO$_3$.
\end{abstract}

\maketitle

\section{Introduction}
Frustration arising from competing interactions gives rise to intriguing noncollinear and noncoplanar magnetic states~\cite{dzyaloshinsky1958thermodynamic,Bogdanov89,jolicoeur1990ground,Okubo_PhysRevLett.108.017206,kitaev2006anyons,batista2016frustration,hayami2021topological}.
Such states are often expressed as a superposition of multiple spin density waves with different wave vectors, which is referred to as multiple-$Q$ states~\cite{yamamoto1972spin,Bak_PhysRevLett.40.800,Shapiro_PhysRevLett.43.1748,bak1980theory,Forgan_PhysRevLett.62.470,batista2016frustration,hayami2021topological}.
The spin at site $j$, $\bm{S}_j$, is generally represented by
\begin{align} 
\label{eq:multiple-q}
\bm{S}_j=\sum_{\eta=1}^n (\bm{S}_{\bm{Q}_\eta}e^{i\bm{Q}_\eta\cdot\bm{R}_j}+\bm{S}_{-\bm{Q}_\eta}e^{-i\bm{Q}_\eta\cdot\bm{R}_j} ),
\end{align}
where $\bm{S}_{\bm{Q}_\eta}$ is the Fourier expansion coefficient of the component at the wave vector $\bm{Q}_\eta$; $\bm{R}_j$ represents the position vector at site $j$.
When $\bm{S}_j$ is mainly characterized by $n=2$ (3) wave vectors, the state is called the double-$Q$ (triple-$Q$) state.
The spin configuration in Eq.~(\ref{eq:multiple-q}) describes various multiple-$Q$ states according to the spin components $\bm{S}_{\bm{Q}_\eta}=(S^x_{\bm{Q}_\eta}, S^y_{\bm{Q}_\eta}, S^z_{\bm{Q}_\eta})$ and the wave vectors $\bm{Q}_\eta=(Q^x_\eta, Q^y_\eta, Q^z_\eta)$, which are determined by the spin interactions and the lattice geometry.  
Indeed, a plethora of multiple-$Q$ states have been so far observed in the materials under cubic, tetragonal, hexagonal, and trigonal lattice structures~\cite{Tokura_doi:10.1021/acs.chemrev.0c00297}.
In the case of the cubic symmetry, the examples are a double-$Q$ state in CeAl$_2$~\cite{forgan1990magnetic}, double-$Q$ meron-antimeron lattice in Co$_8$Zn$_9$Mn$_3$~\cite{yu2018transformation}, triple-$Q$ skyrmion lattice (SkL) in MnSi~\cite{Muhlbauer_2009skyrmion}, triple-$Q$ hedgehog lattice (HL) in MnGe~\cite{Kanazawa_PhysRevLett.106.156603,tanigaki2015real}, quadraple-$Q$ HL in MnSi$_{1-x}$Ge$_x$~\cite{fujishiro2019topological} and SrFeO$_3$~\cite{Ishiwata_PhysRevB.84.054427,Ishiwata_PhysRevB.101.134406}, and triple-$Q$ fractional antiferromagnetic SkL
  in MnSc$_2$S$_4$~\cite{Gao2016Spiral,gao2020fractional}.

The stabilization mechanisms for these multiple-$Q$ states in cubic systems have been theoretically studied based on competing isotropic exchange interactions~\cite{yamamoto1972spin,PhysRevB.74.134411,Okubo_PhysRevB.84.144432, Wang_PhysRevLett.115.107201, PhysRevResearch.2.043278,Aoyama_PhysRevB.103.014406}, anisotropic exchange interactions~\cite{Binz_PhysRevB.74.214408,Park_PhysRevB.83.184406, PhysRevB.88.195137,Yang2016, Okumura_PhysRevB.101.144416,Shimizu_PhysRevB.103.054427,hayami2021field,Mendive-Tapia_PhysRevB.103.024410,kato2022magnetic,kato2021spin,PhysRevB.105.224402}, four(six)-spin interaction~\cite{Okumura_PhysRevB.101.144416,grytsiuk2020topological,Shimizu_PhysRevB.103.054427,hayami2021field,Mendive-Tapia_PhysRevB.103.024410,doi:10.7566/JPSJ.91.093702}, and indirect interactions mediated by itinerant electrons~\cite{jo1984multiple,hirai1985triple,Alonso_PhysRevB.64.054408, Chern_PhysRevLett.105.226403, Hayami_PhysRevB.89.085124,Hayami_PhysRevB.90.060402,kakehashi2018multiple,kakehashi2020numerical,kakehashi2020analytic}.
In particular, the mechanism based on the Dzyaloshinskii-Moriya (DM) interaction~\cite{dzyaloshinsky1958thermodynamic,moriya1960anisotropic}, which is categorized into antisymmetric anisotropic exchange interactions in noncentrosymmetric lattices, has succeeded in explaining various experimental results~\cite{PhysRevB.88.195137}. 
In this case, the appearance of the multiple-$Q$ states is naturally accounted for by Lifshitz invariants in the free energy~\cite{Bogdanov89,rossler2006spontaneous}.
Meanwhile, recent studies have revealed that the symmetric anisotropic exchange interactions, which arise irrespective of the inversion symmetry of the lattice structure, also become the origin of the multiple-$Q$ states in various lattice systems including not only hexagonal~\cite{Hayami_PhysRevB.103.054422,Hirschberger_10.1088/1367-2630/abdef9,utesov2021mean,PhysRevB.106.174437}, trigonal~\cite{yambe2021skyrmion,amoroso2020spontaneous,amoroso2021tuning}, and tetragonal~\cite{Hayami_PhysRevB.103.024439,Wang_PhysRevB.103.104408,Hayami_doi:10.7566/JPSJ.89.103702,doi:10.7566/JPSJ.91.023705,Utesov_PhysRevB.103.064414} systems but also cubic systems~\cite{kato2022magnetic,PhysRevB.105.224402}.
Furthermore, this type of the interactions can lead to different multiple-$Q$ instabilities from those by the DM interaction.
Thus, it is desired to systematically investigate the role of the symmetric anisotropic interactions as well as the antisymmetric ones in cubic systems in order to further explore exotic three-dimensional multiple-$Q$ states.

In this study, we classify both symmetric and antisymmetric exchange interactions according to the cubic symmetry and construct a general anisotropic spin model to examine multiple-$Q$ instabilities in cubic systems.
The obtained model consists of momentum-dependent anisotropic exchange interactions, which is used as a mean-field spin model for insulating magnets or an effective spin model for itinerant magnets with strong Fermi surface nesting~\cite{Hayami_PhysRevLett.121.137202,Okada_PhysRevB.98.224406,PhysRevB.106.174437}.
Following a symmetry argument in Ref.~\cite{PhysRevB.106.174437}, we present the model for the centrosymmetric space groups, $Pm\bar{3}m$ and $Pm\bar{3}$, and the noncentrosymmetric ones, $P\bar{4}3m$, $P432$, and $P23$ in Sec.~\ref{sec:model}. 
As the spin model in each cubic space group has different anisotropic exchange interactions, different multiple-$Q$ instabilities are expected.  
As an example, we show that double-$Q$ and triple-$Q$ states are stabilized by taking into account anisotropic exchange interactions under the $Pm\bar{3}$ symmetry even without an external magnetic field through simulated annealing in Sec.~\ref{sec:simulation}.  
We summarize our results in Sec.~\ref{sec:summary}.
In Appendix~\ref{sec:other_int}, we show the details of the models.

\section{General anisotropic spin model}
\label{sec:model}

\begin{table}
\caption{
\label{tab:rule}
Symmetry rules for nonzero coupling constants given in Ref.~\cite{PhysRevB.106.174437}.
$I$, $m$, and $C_n$ stand for the space inversion, mirror, and $n$-fold rotation operations, respectively (see the text in detail).
The direction of $x_s$ is set along $\bm{q}$.
$\parallel\sharp$ ($\perp\sharp$) for $\sharp=$ plane and axis represents nonzero components parallel (perpendicular) to $\sharp$. 
$-$ means no symmetry constraint.
}
\begin{ruledtabular}
\begin{tabular}{cccc}
symmetry & $\bm{D}_{\bm{q}}$ & $\bm{E}_{\bm{q}}$ & $\bm{F}_{\bm{q}}$ \\ \hline
$I$ & $=\bm{0}$ & $-$ & $-$ \\ 
$m_{\perp}$ & $\parallel$ plane & $\perp$ plane & $-$  \\
$C_{2\perp}$ &  $\perp$ axis  & $\parallel$ axis & $-$ \\
$m_{\parallel}$ & $\perp$ plane & $\perp$ plane & $-$ \\
$C_{2\parallel}$ & $\parallel$ axis & $\parallel$ axis & $-$ \\
$C_{n\parallel}$ ($n \geq 3$) &  $\parallel$ axis & $=\bm{0}$ & $(F^{x_s},F^\perp,F^\perp)$ 
\end{tabular}
\end{ruledtabular}
\end{table}

We consider a general bilinear exchange interaction in momentum space, which is given by
\begin{align}
\label{eq:Interaction}
\bm{S}_{\bm{q}}^T X_{\bm{q}}  \bm{S}_{-\bm{q}},
\end{align}
with
\begin{align}
\label{eq:CouplingMatrix}
X_{\bm{q}}=
\begin{pmatrix}
F_{\bm{q}}^{x_\mathrm{s}} & E_{\bm{q}}^{z_\mathrm{s}}+iD_{\bm{q}}^{z_\mathrm{s}} & E_{\bm{q}}^{y_\mathrm{s}}-iD_{\bm{q}}^{y_\mathrm{s}} \\
E_{\bm{q}}^{z_\mathrm{s}}-iD_{\bm{q}}^{z_\mathrm{s}} & F_{\bm{q}}^{y_\mathrm{s}}  & E_{\bm{q}}^{x_\mathrm{s}}+iD_{\bm{q}}^{x_\mathrm{s}} \\
E_{\bm{q}}^{y_\mathrm{s}}+iD_{\bm{q}}^{y_\mathrm{s}} & E_{\bm{q}}^{x_\mathrm{s}}-iD_{\bm{q}}^{x_\mathrm{s}}  & F_{\bm{q}}^{z_\mathrm{s}}  
\end{pmatrix}.
\end{align}
Here, $\bm{S}_{\bm{q}}=(S^{x_\mathrm{s}}_{\bm{q}},S^{y_\mathrm{s}}_{\bm{q}},S^{z_\mathrm{s}}_{\bm{q}})$ is the Fourier transform of the spin, $(x_\mathrm{s}, y_\mathrm{s}, z_\mathrm{s})$ are Cartesian spin coordinates, and $T$ denotes the transpose of the vector.
$X_{\bm{q}}$ represents the general interaction matrix in spin space, which consists of three types of real coupling constants $\bm{D}_{\bm{q}}=(D^{x_\mathrm{s}}_{\bm{q}}, D^{y_\mathrm{s}}_{\bm{q}}, D^{z_\mathrm{s}}_{\bm{q}})$ for DM-type antisymmetric interactions, $\bm{E}_{\bm{q}}=(E^{x_\mathrm{s}}_{\bm{q}}, E^{y_\mathrm{s}}_{\bm{q}}, E^{z_\mathrm{s}}_{\bm{q}})$ for off-diagonal symmetric interactions, and $\bm{F}_{\bm{q}}=(F^{x_\mathrm{s}}_{\bm{q}}, F^{y_\mathrm{s}}_{\bm{q}}, F^{z_\mathrm{s}}_{\bm{q}})$ for diagonal symmetric interactions.
The interaction matrix satisfies $X_{\bm{q}}^*=X_{-\bm{q}}$; 
$\bm{D}_{\bm{q}}=-\bm{D}_{-\bm{q}}$, $\bm{E}_{\bm{q}}=\bm{E}_{-\bm{q}}$, and $\bm{F}_{\bm{q}}=\bm{F}_{-\bm{q}}$.
We neglect the sublattice degree of freedom in this paper, while its extension is straightforwardly applied in the same manner as Ref.~\cite{PhysRevB.106.174437}.

The interaction in Eq.~(\ref{eq:Interaction}) is defined on the ``bond" between the wave vectors $\pm\bm{q}$, which means that nonzero components in $X_{\bm{q}}$ are determined according to the transformation in terms of point group symmetries leaving the bond:
space inversion ($I$), mirror perpendicular to $\bm{q}$ ($m_\perp$), twofold rotation perpendicular to $\bm{q}$ ($C_{2\perp}$), mirror parallel to $\bm{q}$ ($m_\parallel$), and $n$-fold ($n=2,3,4,6$) rotation around $\bm{q}$ ($C_{n\parallel}$).
The symmetry rules for nonzero coupling constants were obtained by using magnetic representation theory in Ref.~\cite{PhysRevB.106.174437}, which is summarized in Table~\ref{tab:rule}. 
By applying these rules for the cubic space groups, one can obtain nonzero coupling constants in each wave vector in the Brillouin zone.
These symmetry rules are applicable to all the wave vectors except for the time-reversal invariant wave vectors at the Brillouin zone boundary, where $\bm{D}_{\bm{q}}=\bm{0}$ irrespective of the inversion symmetry.
It is noted that there are additional constraints between the interactions at $\bm{q}$ and $\bm{q}'\neq\pm\bm{q}$, once the rotational symmetry of the cubic systems is taken into account.
For example, the interaction components at $\bm{q}=(q,0,0)$ are related to those at $\bm{q}'=(0,q,0)$ and $\bm{q}''=(0,0,q)$ under threefold rotational symmetry around the [111] axis, as discussed in Appendix~\ref{sec:other_int}.

\begin{figure}[t!]
\begin{center}
\includegraphics[width=1.0\hsize]{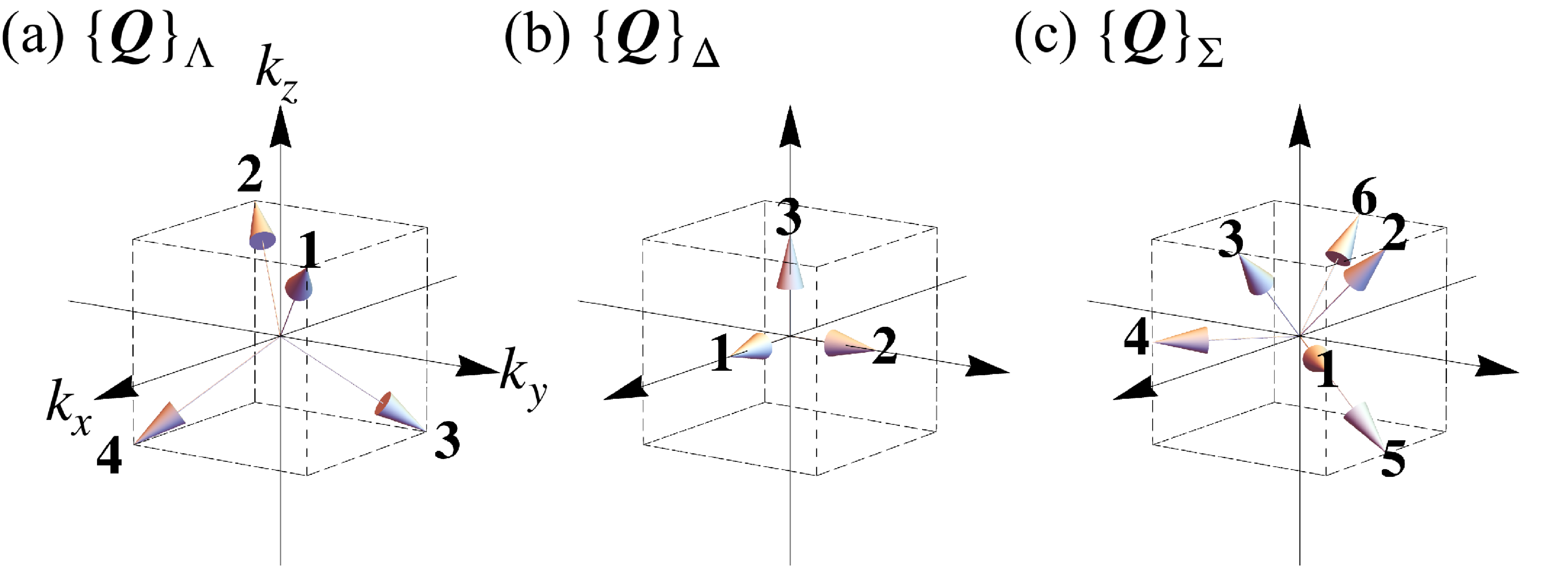} 
\caption{
\label{fig:BZ} 
High-symmetry wave vectors in cubic systems: (a) $\{\bm{Q}\}_\Lambda\ni\bm{Q}_1=(Q,Q,Q)$, $\bm{Q}_2=(-Q,-Q,Q)$, $\bm{Q}_3=(-Q,Q,-Q)$, and $\bm{Q}_4=(Q,-Q,-Q)$, (b) $\{\bm{Q}\}_\Delta\ni\bm{Q}_1=(Q,0,0)$, $\bm{Q}_2=(0,Q,0)$, and $\bm{Q}_3=(0,0,Q)$, and (c) $\{\bm{Q}\}_\Sigma\ni\bm{Q}_1=(Q,Q,0)$, $\bm{Q}_2=(0,Q,Q)$, $\bm{Q}_3=(Q,0,Q)$, $\bm{Q}_4=(Q,-Q,0)$, $\bm{Q}_5=(0,Q,-Q)$, and $\bm{Q}_6=(-Q,0,Q)$.
The wave vector $\bm{Q}_\eta$ is shown by the arrow labeled by $\eta=1$--$6$.
}
\end{center}
\end{figure}

\begin{table*}
\caption{\label{tab:high_symmetry}
Interaction matrix $X_{\bm{Q}_1}$ and the number of independent components $N_\mathrm{c}$ in the cubic systems for the high-symmetry wave vectors shown in Fig.~\ref{fig:BZ}: $\bm{Q}_1\parallel[111]\in\{\bm{Q}\}_\Lambda$, $\bm{Q}_1\parallel[100]\in\{\bm{Q}\}_\Delta$, and $\bm{Q}_1\parallel[110]\in\{\bm{Q}\}_\Sigma$.
The spin coordinates $x_\mathrm{s}$, $y_\mathrm{s}$, and $z_\mathrm{s}$ are taken along the $x$, $y$, and $z$ directions, respectively.
The checkmark ($\checkmark$) shows the presence of the inversion symmetry $I$. 
}
\begin{ruledtabular}
\begin{tabular}{ccccccc}
   &\multicolumn{2}{c}{$\bm{Q}_1\parallel[111]\in\{\bm{Q}\}_\Lambda$}  & \multicolumn{2}{c}{$\bm{Q}_1\parallel[100]\in\{\bm{Q}\}_\Delta$}   & \multicolumn{2}{c}{$\bm{Q}_1\parallel[110]\in\{\bm{Q}\}_\Sigma$}  \\
 \cline{2-3} \cline{4-5}  \cline{6-7}
space group & $X_{\bm{Q}_1}$ & $N_\mathrm{c}$ & $X_{\bm{Q}_1}$ & $N_\mathrm{c}$  & $X_{\bm{Q}_1}$ & $N_\mathrm{c}$ 
 \\ \hline
$Pm\bar{3}m$ ($\checkmark$) 
& $\begin{pmatrix} F_{\bm{Q}_1}^{x} & E_{\bm{Q}_1}^{x} & E_{\bm{Q}_1}^{x} \\ E_{\bm{Q}_1}^{x} & F_{\bm{Q}_1}^{x}  & E_{\bm{Q}_1}^{x} \\ E_{\bm{Q}_1}^{x} & E_{\bm{Q}_1}^{x}  & F_{\bm{Q}_1}^{x}  \end{pmatrix}$ & 2
& $\begin{pmatrix} F_{\bm{Q}_1}^{x} & 0 & 0 \\ 0 & F_{\bm{Q}_1}^{y}  & 0 \\ 0& 0  & F_{\bm{Q}_1}^{y}  \end{pmatrix}$ & 2 
& $\begin{pmatrix} F_{\bm{Q}_1}^{x} & E_{\bm{Q}_1}^{z} & 0 \\ E_{\bm{Q}_1}^{z} & F_{\bm{Q}_1}^{x}  & 0 \\ 0& 0  & F_{\bm{Q}_1}^{z}  \end{pmatrix}$ & 3
  \\
$P\bar{4}3m$
& $\begin{pmatrix} F_{\bm{Q}_1}^{x} & E_{\bm{Q}_1}^{x} & E_{\bm{Q}_1}^{x} \\ E_{\bm{Q}_1}^{x} & F_{\bm{Q}_1}^{x}  & E_{\bm{Q}_1}^{x} \\ E_{\bm{Q}_1}^{x} & E_{\bm{Q}_1}^{x}  & F_{\bm{Q}_1}^{x}  \end{pmatrix}$ & 2 
& $\begin{pmatrix} F_{\bm{Q}_1}^{x} & 0 & 0 \\ 0 & F_{\bm{Q}_1}^{y}  & 0 \\ 0& 0  & F_{\bm{Q}_1}^{y}  \end{pmatrix}$ & 2
& $\begin{pmatrix} F_{\bm{Q}_1}^{x} & E_{\bm{Q}_1}^{z} & iD_{\bm{Q}_1}^{x} \\ E_{\bm{Q}_1}^{z} & F_{\bm{Q}_1}^{x}  &  iD_{\bm{Q}_1}^{x} \\ - iD_{\bm{Q}_1}^{x}& - iD_{\bm{Q}_1}^{x}  & F_{\bm{Q}_1}^{z}  \end{pmatrix}$ & 4  
\\
$P432$ 
& $\begin{pmatrix} F_{\bm{Q}_1}^{x} & E_{\bm{Q}_1}^{x}+iD_{\bm{Q}_1}^{x} & E_{\bm{Q}_1}^{x}-iD_{\bm{Q}_1}^{x} \\ E_{\bm{Q}_1}^{x}-iD_{\bm{Q}_1}^{x} & F_{\bm{Q}_1}^{x}  & E_{\bm{Q}_1}^{x}+iD_{\bm{Q}_1}^{x} \\ E_{\bm{Q}_1}^{x}+iD_{\bm{Q}_1}^{x} & E_{\bm{Q}_1}^{x}-iD_{\bm{Q}_1}^{x}  & F_{\bm{Q}_1}^{x}  \end{pmatrix}$ & 3
& $\begin{pmatrix} F_{\bm{Q}_1}^{x} & 0 & 0 \\ 0 & F_{\bm{Q}_1}^{y}  & iD_{\bm{Q}_1}^{x} \\ 0& -iD_{\bm{Q}_1}^{x}  & F_{\bm{Q}_1}^{y}  \end{pmatrix}$ & 3  
& $\begin{pmatrix} F_{\bm{Q}_1}^{x} & E_{\bm{Q}_1}^{z} & -iD_{\bm{Q}_1}^{x} \\ E_{\bm{Q}_1}^{z} & F_{\bm{Q}_1}^{x}  &  iD_{\bm{Q}_1}^{x} \\ iD_{\bm{Q}_1}^{x}& - iD_{\bm{Q}_1}^{x}  & F_{\bm{Q}_1}^{z}  \end{pmatrix}$ & 4  
 \\
 $Pm\bar{3}$ ($\checkmark$) 
 & $\begin{pmatrix} F_{\bm{Q}_1}^{x} & E_{\bm{Q}_1}^{x} & E_{\bm{Q}_1}^{x} \\ E_{\bm{Q}_1}^{x} & F_{\bm{Q}_1}^{x}  & E_{\bm{Q}_1}^{x} \\ E_{\bm{Q}_1}^{x} & E_{\bm{Q}_1}^{x}  & F_{\bm{Q}_1}^{x}  \end{pmatrix}$ & 2  
 & $\begin{pmatrix} F_{\bm{Q}_1}^{x} & 0 & 0 \\ 0 & F_{\bm{Q}_1}^{y}  & 0 \\ 0& 0  & F_{\bm{Q}_1}^{z}  \end{pmatrix}$ & 3 
 & $\begin{pmatrix} F_{\bm{Q}_1}^{x} & E_{\bm{Q}_1}^{z} & 0 \\ E_{\bm{Q}_1}^{z} & F_{\bm{Q}_1}^{y}  & 0 \\ 0& 0  & F_{\bm{Q}_1}^{z}  \end{pmatrix}$ & 4 
 \\
$P23$ 
& $\begin{pmatrix} F_{\bm{Q}_1}^{x} & E_{\bm{Q}_1}^{x}+iD_{\bm{Q}_1}^{x} & E_{\bm{Q}_1}^{x}-iD_{\bm{Q}_1}^{x} \\ E_{\bm{Q}_1}^{x}-iD_{\bm{Q}_1}^{x} & F_{\bm{Q}_1}^{x}  & E_{\bm{Q}_1}^{x}+iD_{\bm{Q}_1}^{x} \\ E_{\bm{Q}_1}^{x}+iD_{\bm{Q}_1}^{x} & E_{\bm{Q}_1}^{x}-iD_{\bm{Q}_1}^{x}  & F_{\bm{Q}_1}^{x}  \end{pmatrix}$ & 3 
& $\begin{pmatrix} F_{\bm{Q}_1}^{x} & 0 & 0 \\ 0 & F_{\bm{Q}_1}^{y}  & iD_{\bm{Q}_1}^{x} \\ 0& -iD_{\bm{Q}_1}^{x}  & F_{\bm{Q}_1}^{z}  \end{pmatrix}$ & 4
& $\begin{pmatrix} F_{\bm{Q}_1}^{x} & E_{\bm{Q}_1}^{z} & -iD_{\bm{Q}_1}^{y} \\ E_{\bm{Q}_1}^{z} & F_{\bm{Q}_1}^{y}  &  iD_{\bm{Q}_1}^{x} \\ iD_{\bm{Q}_1}^{y}& - iD_{\bm{Q}_1}^{x}  & F_{\bm{Q}_1}^{z}  \end{pmatrix}$ & 6 
\end{tabular}
\end{ruledtabular}
\end{table*}

When considering the magnetic instability from high temperatures or at low temperatures close to the ground state, it is enough to consider the dominant interaction channels at specific wave vectors in momentum space in determining the optimal spin configuration from the energetic point of view. 
Based on this consideration, we construct an anisotropic spin model consisting of specific wave-vector interactions, which is given by
\begin{align}
\label{eq:model}
\mathcal{H}
= -\sum_{\bm{q}\in\{\bm{Q}\}
}\bm{S}_{\bm{q}}^T X_{\bm{q}}  \bm{S}_{-\bm{q}}.
\end{align}
Here, $\{\bm{Q}\}=\{\bm{Q}_1,\bm{Q}_2,\cdots,\bm{Q}_n\}$ is a set of the symmetry-related wave vectors, and $\bm{q}\in\{\bm{Q}\}$ gives the largest eigenvalue of $X_{\bm{q}}$.
The model has, at most, nine independent parameters, since the interactions at $\{\bm{Q}\}$ are related to each other under point group symmetry.
In other words, the interaction parameters in $X_{\bm{Q}_{\eta\neq1}}$ are expressed as those in $X_{\bm{Q}_1}$. 
Thus, it is enough to obtain $X_{\bm{Q}_1}$ in each space group.
We show the results under five cubic space groups, $Pm\bar{3}m$, $P\bar{4}3m$, $P432$, $Pm\bar{3}$, and $P23$ in Table~\ref{tab:high_symmetry}.
In each space group, we present interaction matrices with three different high-symmetry $\{\bm{Q}\}$:
$\{\bm{Q}\}_\Lambda\ni\bm{Q}_1\parallel[111]$ shown in Fig.~\ref{fig:BZ}(a).
$\{\bm{Q}\}_\Delta\ni\bm{Q}_1\parallel[100]$ shown in Fig.~\ref{fig:BZ}(b), 
and $\{\bm{Q}\}_\Sigma\ni\bm{Q}_1\parallel[110]$ shown in Fig.~\ref{fig:BZ}(c). 
We also present the number of independent coupling constants $N_\mathrm{c}\ge2$, which includes the isotropic interaction $F^\mathrm{iso}_{\bm{Q}_1}=(F^x_{\bm{Q}_1}+F^y_{\bm{Q}_1}+F^z_{\bm{Q}_1})/3$ appearing irrespective of the space group and wave vector.
The remaining interactions at $\bm{Q}_{\eta\neq1}$ are shown in Appendix~\ref{sec:other_int}.
  
We discuss the similarity and difference of $X_{\bm{Q}_1}$ between five space groups in each high-symmetry wave vector. 
In the case of $\bm{Q}_1\parallel[111]\in\{\bm{Q}\}_\Lambda$ shown in the left column in Table~\ref{tab:high_symmetry}, there are at least two independent coupling constants ($N_\mathrm{c}\ge2$) in $X_{\bm{Q}_1}$ irrespective of the cubic space groups: One is the isotropic interaction $F^\mathrm{iso}_{\bm{Q}_1}$ and the other is the uniaxially anisotropic interaction $E^x_{\bm{Q}_1}$ along the $\bm{Q}_1$ direction, the latter of which arises from the symmetry rule in terms of $C_{3\parallel}$. 
The positive (negative) anisotropic interaction $E^x_{\bm{Q}_1}$ corresponds to the easy-axis (easy-plane) interaction along the [111] direction, which favors the spin modulation parallel (perpendicular) to $\bm{Q}_1$. 
In addition, the DM interaction $D^x_{\bm{Q}_1}$ appears in noncentrosymmetric space groups $P432$ and $P23$, which favors the proper-screw spiral modulation on the plane perpendicular to $\bm{Q}_1$. 
Meanwhile, there is no DM interaction in the other noncentrosymmetric space group $P\bar{4}3m$ due to the presence of $m_\parallel$ on the plane perpendicular to [1$\bar{1}$0].  
Thus, the multiple-$Q$ instability in the $P\bar{4}3m$ system is qualitatively similar to that in the centrosymmetric $Pm\bar{3}m$ and $Pm\bar{3}$ systems rather than the noncentrosymmetric $P432$ and $P23$ systems.
 
The result for $\bm{Q}_1\parallel[100]\in\{\bm{Q}\}_\Delta$ is shown in the middle column of Table~\ref{tab:high_symmetry}. 
Similar to the case of $\bm{Q}_1 \parallel [111]$, the interaction matrices are characterized by at least two independent coupling constants ($N_\mathrm{c}\ge 2$).
The difference from the result for $\bm{Q}_1 \parallel [111]$ appears in the easy-axis direction of the uniaxially anisotropic interaction; $F^x_{\bm{Q}_1}>F^y_{\bm{Q}_1}$ ($F^x_{\bm{Q}_1}<F^y_{\bm{Q}_1}$) corresponds to the easy-axis (easy-plane) interaction to favor the spin modulation parallel (perpendicular) to $\bm{Q}_1\parallel[100]$. 
The interaction matrix for $Pm\bar{3}m$ is characterized by these two components.
In addition, the interaction matrix for $P\bar{4}3m$ also has the same two independent components in spite of the noncentrosymmetric lattice structure; two symmetry rules in terms of $C_{2\parallel}$ and $m_\parallel$ on the plane perpendicular to [01$\bar{1}$] axis impose on no additional component. 
The DM interaction appears in the interaction matrix for $P432$ and $P23$, which tends to favor the proper-screw spiral modulation. 
Furthermore, the additional symmetric exchange interaction in $\bm{F}_{\bm{Q}_1}$ appears for $Pm\bar{3}$ and $P23$.
The relation with $F^x_{\bm{Q}_1}\neq F^y_{\bm{Q}_1}\neq F^z_{\bm{Q}_1}$ is owing to a triaxial anisotropy in the absence of fourfold rotational symmetry around the [100] axis. 

The result for $\bm{Q}_1\parallel[110]\in\{\bm{Q}\}_\Sigma$ is presented in the right column in Table~\ref{tab:high_symmetry}.
Compared to the [111] and [100] directions, the number of independent components increases.
There are at least three independent coupling constants ($N_\mathrm{c}\ge 3$).
The interaction matrix for $Pm\bar{3}m$ ($Pm\bar{3}$) is characterized by the triaxially anisotropic interaction with independent $F^x_{\bm{Q}_1}$, $F^y_{\bm{Q}_1}$, and $F^z_{\bm{Q}_1}$ ($F^x_{\bm{Q}_1}$, $F^y_{\bm{Q}_1}$, $F^z_{\bm{Q}_1}$, and $E^z_{\bm{Q}_1}$).
The interaction matrices for $P\bar{4}3m$ and $P432$ also have the triaxial anisotropy with $F^x_{\bm{Q}_1}$, $F^z_{\bm{Q}_1}$, and $E^z_{\bm{Q}_1}$ in the symmetric component. Besides, these space groups exhibit the antisymmetric component $D^x_{\bm{Q}_1}$. 
The DM vector lies on the plane parallel (perpendicular) to $\bm{Q}_1$ for $P432$ ($P\bar{4}3m$), which tends to favor the proper-screw (cycloidal) spiral modulation.
In contrast to the cases of $\bm{Q}_1 \parallel [111]$ and $\bm{Q}_1 \parallel [001]$, the DM interaction appears in the $P\bar{4}3m$ system for $\bm{Q}_1 \parallel [110]$, which can become the origin of the multiple-$Q$ states.
The interaction matrix for $P23m$ is expressed as the triaxial symmetric anisotropic interactions with $F^x_{\bm{Q}_1}$, $F^y_{\bm{Q}_1}$, $F^z_{\bm{Q}_1}$, and $E^z_{\bm{Q}_1}$ and the DM interactions with $D^x_{\bm{Q}_1}$ and $D^y_{\bm{Q}_1}$.
In this case, the spiral plane lies on the plane neither parallel nor perpendicular to $\bm{Q}_1$.

The anisotropic spin model in Eq.~(\ref{eq:model}) was used to investigate the multiple-$Q$ instabilities in noncentrosymmetric cubic systems~\cite{Okumura_PhysRevB.101.144416,Shimizu_PhysRevB.103.054427,hayami2021field,kato2022magnetic}.
In particular, the models in Ref.~\cite{kato2022magnetic} are exactly the same as those for $P23$ in Table~\ref{tab:high_symmetry}.
These previous studies showed that the DM interaction combined with the symmetric anisotropic interaction~\cite{kato2022magnetic}, four-spin interaction~\cite{Okumura_PhysRevB.101.144416,Shimizu_PhysRevB.103.054427,hayami2021field}, or magnetic field~\cite{hayami2021field} stabilizes the multiple-$Q$ states in the ground state, and discuss the origin of the HL in MnSi$_{1-x}$Ge$_x$~\cite{Kanazawa_PhysRevLett.106.156603,tanigaki2015real,fujishiro2019topological} and the SkL in EuPtSi~\cite{kakihana2018giant,kaneko2019unique,kakihana2019unique,tabata2019magnetic}.
Meanwhile, the multiple-$Q$ instabilities have not been studied in the anisotropic spin models in centrosymmetric cubic systems, which we analyze in Sec.~\ref{sec:simulation}.   

\section{Simulation result}
\label{sec:simulation}

To demonstrate that the anisotropic spin model gives rise to a variety of multiple-$Q$ states,
we numerically analyze the model at $\{\bm{Q}\}_\Delta$ on a simple cubic lattice under the space group $Pm\bar{3}$, which is given by
\begin{align}
\label{eq:model2}
\mathcal{H}
= -2\sum_{\bm{q}\in\{\bm{Q}_\Delta\}
}\bm{S}_{\bm{q}}^T X_{\bm{q}}  \bm{S}_{-\bm{q}},
\end{align}
where
\begin{align}
X_{\bm{Q}_1}&=\begin{pmatrix}
F_{\bm{Q}_1}^{x} & 0 & 0\\
0 & F_{\bm{Q}_1}^{y}  & 0 \\
0 & 0  & F_{\bm{Q}_1}^{z}
\end{pmatrix},\\
X_{\bm{Q}_2}&=\begin{pmatrix}
F_{\bm{Q}_1}^{z} & 0 & 0\\
0 & F_{\bm{Q}_1}^{x}  & 0 \\
0 & 0  & F_{\bm{Q}_1}^{y}
\end{pmatrix},\\
X_{\bm{Q}_3}&=\begin{pmatrix}
F_{\bm{Q}_1}^{y} & 0 & 0 \\
0 & F_{\bm{Q}_1}^{z}  & 0 \\
0& 0 & F_{\bm{Q}_1}^{x}
\end{pmatrix},
\end{align}
and $\bm{Q}_1=(Q, 0, 0)$, $\bm{Q}_{2}=(0, Q, 0)$, and $\bm{Q}_{3}=(0, 0, Q)$ with $Q=\pi/3$; the lattice constant of the cubic lattice is taken as unity.
The interaction matrices at $\bm{Q}_2$ and $\bm{Q}_3$ are expressed as $\bm{F_{\bm{Q}_1}}$ due to the threefold rotation along the [111] axis (see Appendix~\ref{sec:other_int}). 
The coefficient 2 in Eq.~(\ref{eq:model2}) is introduced to take into account the interaction at $-\bm{Q}_\eta$.
We fix the spin length at each site as unity for simplicity. 
It is noted that the model in Eq.~(\ref{eq:model2}) corresponds to that in the $Pm\bar{3}m$ and $P\bar{4}3m$ systems by setting $F_{\bm{Q}_1}^{y}=F_{\bm{Q}_1}^{z}$.

The ground-state phase diagram is calculated by simulated annealing combined with the standard Metropolis local updates in real space.
Starting from a high temperature $T_0$, we gradually reduce the temperature with a rate $T_{n+1}=\alpha T_n$ to a final temperature $T_\mathrm{f}=0.01$, where $T_n$ is the temperature at the $n$th step.
Typically, we set $T_0=1$--$10$ and $\alpha=0.999995$, and we spend around $10^{6}$ Monte Carlo steps for annealing.
After reaching the final temperature, we perform  $10^{6}$ Monte Carlo steps for thermalization and measurements, respectively.
To identify magnetic phases, we calculate a spin structure factor given by
\begin{align}
S^\alpha_s (\bm{q})=\left\langle
\frac{1}{N}\sum_{j,k}S^\alpha_jS^\alpha_k e^{i\bm{q}\cdot(\bm{R}_j-\bm{R}_k)}
\right\rangle,
\end{align}
where $\alpha=x,y,z$, $N$ is the system size, $\bm{S}_j$ is the classical spin at site $j$ ($|\bm{S}_j|=1$), $\bm{R}_j$ is the position vector, and $\langle\cdots\rangle$ is the average over the Monte Carlo samples.
In the following, we show the result for $N=12^3$ under the periodic boundary conditions.

\begin{figure}[t!]
\begin{center}
\includegraphics[width=1.0\hsize]{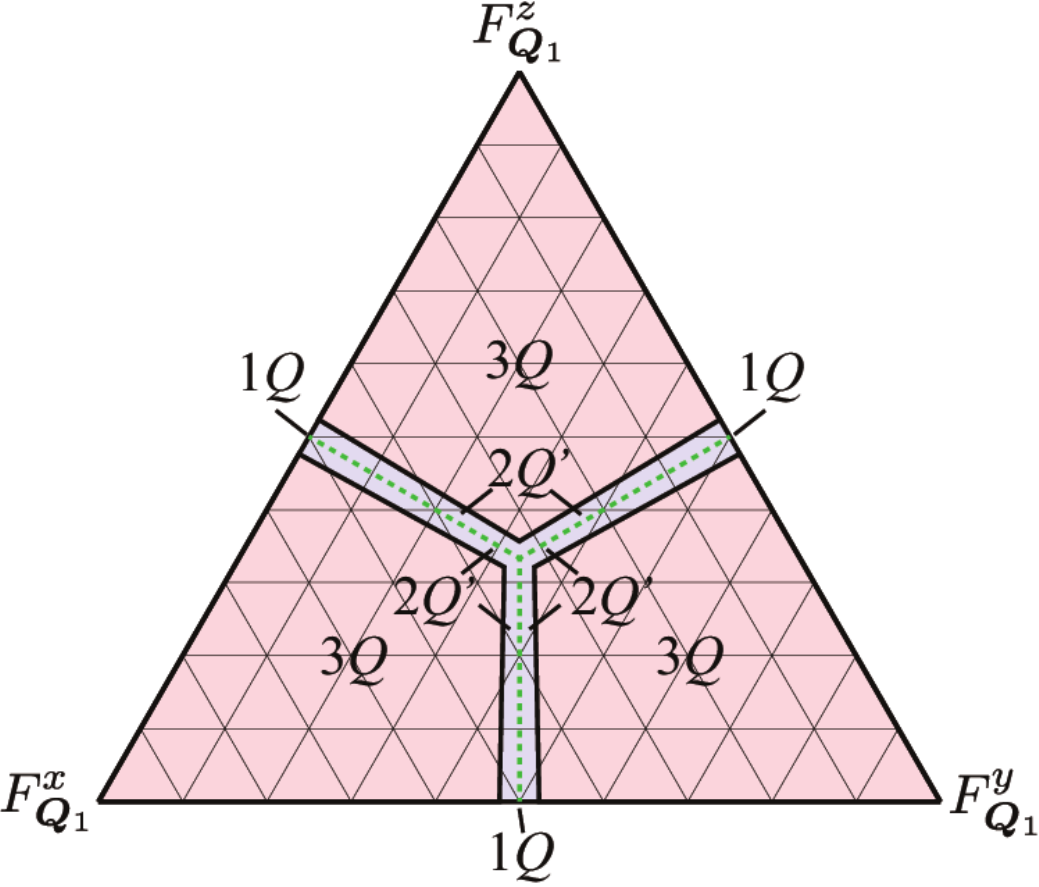} 
\caption{
\label{fig:PD} 
Magnetic phase diagram under $F^x_{\bm{Q}_1}+F^y_{\bm{Q}_1}+F^z_{\bm{Q}_1}=1$.
The dashed green lines represent the region where the 1$Q$ state is stabilized. 
}
\end{center}
\end{figure}

We show the ground-state phase diagram in Fig.~\ref{fig:PD}, where $F^x_{\bm{Q}_1}+F^y_{\bm{Q}_1}+F^z_{\bm{Q}_1}=1$ and  $F^\alpha_{\bm{Q}_1}\ge0$.
The phase diagram is threefold symmetric in terms of the point at $F^x_{\bm{Q}_1}=F^y_{\bm{Q}_1}=F^z_{\bm{Q}_1}$ and twofold symmetric in terms of the lines at $F^x_{\bm{Q}_1} \neq F^y_{\bm{Q}_1}=F^z_{\bm{Q}_1}$, $F^y_{\bm{Q}_1} \neq F^z_{\bm{Q}_1}=F^x_{\bm{Q}_1}$, and $F^z_{\bm{Q}_1} \neq F^x_{\bm{Q}_1}=F^y_{\bm{Q}_1}$.
We find three phases characterized by the single-$Q$ (1$Q$), double-$Q'$ (2$Q'$), and triple-$Q$ (3$Q$) spin configurations in the ground state depending on the interactions; $2Q'$ means the double-$Q$ structure with different intensities at $\{\bm{Q}\}$.
The 1$Q$ state has a coplanar structure, while the 2$Q'$ and 3$Q$ states have noncoplanar ones.
It is noted that these phases on the line at $F^y_{\bm{Q}_1}=F^z_{\bm{Q}_1}$ are also stabilized in $Pm\bar{3}m$ and $P\bar{4}3m$ systems. 

The 1$Q$ state is a spiral state characterized by the single peak of the spin structure factor at $\bm{Q}_1$, $\bm{Q}_2$, or $\bm{Q}_3$. 
This state becomes the ground state when two out of three interaction parameters are the same and they are greater than or equal to the remaining parameter denoted as the green dashed lines in Fig.~\ref{fig:PD}.
In the isotropic case, i.e., $F^x_{\bm{Q}_1}=F^y_{\bm{Q}_1}=F^z_{\bm{Q}_1}$, the spiral plane is arbitrary and irrespective of $\bm{Q}_\eta$.
Meanwhile, in the region for $F^x_{\bm{Q}_1} < F^y_{\bm{Q}_1}=F^z_{\bm{Q}_1}$, $F^y_{\bm{Q}_1} < F^z_{\bm{Q}_1}=F^x_{\bm{Q}_1}$, and $F^z_{\bm{Q}_1} < F^x_{\bm{Q}_1}=F^y_{\bm{Q}_1}$ the spiral plane is fixed depending on $\bm{Q}_\eta$.
For example, the anisotropic interaction with $F^x_{\bm{Q}_1}<F^y_{\bm{Q}_1}=F^z_{\bm{Q}_1}$ fixes the spiral plane on the $yz$ plane at $\bm{Q}_1$, the $zx$ plane at $\bm{Q}_2$, or the $xy$ plane at $\bm{Q}_3$, which are connected by the threefold rotation around the [111] direction.      

When one of the three interaction parameters is slightly greater than the remaining two parameters, the infinitesimal easy-axis anisotropy continuously changes the 1$Q$ state into the 2$Q'$ state, which is characterized by the double peaks of the spin structure factor with different intensities.  
The 2$Q'$ state is expressed as the superposition of the spiral wave and the sinusoidal wave, where the oscillating direction of the sinusoidal wave is perpendicular to the spiral plane~\cite{Ozawa_doi:10.7566/JPSJ.85.103703,yambe2020double,Okumura_PhysRevB.101.144416,Shimizu_PhysRevB.103.054427,kato2022magnetic,doi:10.7566/JPSJ.91.093702}.
For example, the interaction with $F^x_{\bm{Q}_1}>F^y_{\bm{Q}_1} \ge F^z_{\bm{Q}_1}$ stabilizes the 2$Q'$ state with the sinusoidal wave along the $x$ direction at $\bm{Q}_1$ and the spiral wave on the $yz$ plane at $\bm{Q}_2$.
Similar to the $1Q$ state, there are three 2$Q'$ states with the same energy at each parameter due to the threefold rotation around the [111] direction.
The 2$Q$ state has a noncoplanar magnetic structure, where the magnetic vortex and antivortex form the square lattice~\cite{Ozawa_doi:10.7566/JPSJ.85.103703}.

By further increasing the easy-axis anisotropy, the ground state becomes the 3$Q$ state characterized by the triple peaks of the spin structure factor with the same intensity.    
The 3$Q$ state consists of three sinusoidal waves at $\bm{Q}_1$--$\bm{Q}_3$, where the oscillating directions of the sinusoidal waves are orthogonal to each other~\cite{Okumura_PhysRevB.101.144416,Shimizu_PhysRevB.103.054427,kato2022magnetic,doi:10.7566/JPSJ.91.093702}.
For example, in the 3$Q$ state stabilized by $F^x_{\bm{Q}_1}>F^y_{\bm{Q}_1}\ge F^z_{\bm{Q}_1}$, the constitute waves are the sinusoidal waves along the $x$ direction at $\bm{Q}_1$, the $y$ direction at $\bm{Q}_2$, and the $z$ direction at $\bm{Q}_3$.  
A noncoplanar magnetic structure in the 3$Q$ state is regarded as the simple cubic lattice of the magnetic hedgehog and antihedgehog~\cite{kato2022magnetic}.

\section{Summary}
\label{sec:summary}

We present the anisotropic spin model with both the momentum-dependent DM interaction and symmetric anisotropic interaction in cubic systems.
We clarify the nonzero anisotropic interactions at three high-symmetry wave vectors in the $Pm\bar{3}m$, $Pm\bar{3}$, $P\bar{4}3m$, $P432$, and $P23$ cubic space groups based on the symmetry rules.
The results show that the anisotropic interactions largely depend on not only the space group but also the wave vector, which implies that a plethora of multiple-$Q$ states appear by the anisotropic interactions in cubic systems. 
To demonstrate it, we investigate the ground-state phase diagram for centrosymmetric $Pm\bar{3}$ system by simulated annealing. 
We reveal that the symmetric anisotropic interactions stabilize the noncoplanar double-$Q$ and triple-$Q$ states, which are regarded as the vortex-antivortex square lattice and the hedgehog-antihedgehog cubic lattice, respectively.
Our results make it possible to systematically investigate the multiple-$Q$ instability in centrosymmetric and noncentrosymmetric cubic systems based on the anisotropic interactions.    
Such systematic studies will be a good reference for searching new noncollinear and noncoplanar magnetic materials and understanding their origin.

\begin{acknowledgments}
This research was supported by JSPS KAKENHI Grants Numbers JP21H01037, JP22H04468, JP22H00101, JP22H01183, and by JST PRESTO (JPMJPR20L8). 
R.Y. was supported by Forefront Physics and Mathematics Program to Drive Transformation (FoPM).
Parts of the numerical calculations were performed in the supercomputing systems in ISSP, the University of Tokyo.
\end{acknowledgments}

\appendix*

\section{Interactions at the other high-symmetry wave vectors}\label{sec:other_int}

We show the interaction matrices for $Pm\bar{3}m$, $P\bar{4}3m$, $P432$, $Pm\bar{3}$, and $P23$ at $\bm{Q}_2$--$\bm{Q}_{4} $ in $\{\bm{Q}\}_\Lambda$, at $\bm{Q}_2$--$\bm{Q}_3$ in $\{\bm{Q}\}_\Delta$, and at $\bm{Q}_2$--$\bm{Q}_6$ in $\{\bm{Q}\}_\Sigma$ shown in Fig.~\ref{fig:BZ}, whose components are represented by $X_{\bm{Q}_1}$ in Table~\ref{tab:high_symmetry}.
To explicitly obtain nonzero components in the matrices, we use the following point group symmetries: 
the twofold rotation around [100] ($C_{2[100]}$),
twofold rotation around [010] ($C_{2[010]}$),
twofold rotation around [001] ($C_{2[001]}$),
threefold counterclockwise rotation around [111] ($C^+_{3[111]}$),
and threefold clockwise rotation around [111] ($C^-_{3[111]}$).

\subsection{$\{\bm{Q}\}_\Lambda$}

The wave vectors $\bm{Q}_2$, $\bm{Q}_3$, and $\bm{Q}_4$ shown in Fig.~\ref{fig:BZ}(a) are connected to $\bm{Q}_1$ as $\bm{Q}_2=C_{2[001]}\bm{Q}_1$, $\bm{Q}_3=C_{2[010]}\bm{Q}_1$, and $\bm{Q}_4=C_{2[100]}\bm{Q}_1$.   
Then, $X_{\bm{Q}_2}$, $X_{\bm{Q}_3}$, and $X_{\bm{Q}_4}$ for $P432$ and $P23$ are given by
\begin{align}
\label{eq:lambda_2}
X_{\bm{Q}_2}&=\begin{pmatrix}
F_{\bm{Q}_1}^{x} & E_{\bm{Q}_1}^{x}+iD_{\bm{Q}_1}^{x} & -E_{\bm{Q}_1}^{x}+iD_{\bm{Q}_1}^{x} \\
E_{\bm{Q}_1}^{x}-iD_{\bm{Q}_1}^{x} & F_{\bm{Q}_1}^{x}  & -E_{\bm{Q}_1}^{x}-iD_{\bm{Q}_1}^{x} \\
-E_{\bm{Q}_1}^{x}-iD_{\bm{Q}_1}^{x} & -E_{\bm{Q}_1}^{x}+iD_{\bm{Q}_1}^{x}  & F_{\bm{Q}_1}^{x}  
\end{pmatrix},\\
\label{eq:lambda_3}
X_{\bm{Q}_3}&=\begin{pmatrix}
F_{\bm{Q}_1}^{x} & -E_{\bm{Q}_1}^{x}-iD_{\bm{Q}_1}^{x} & E_{\bm{Q}_1}^{x}-iD_{\bm{Q}_1}^{x} \\
-E_{\bm{Q}_1}^{x}+iD_{\bm{Q}_1}^{x} & F_{\bm{Q}_1}^{x}  & -E_{\bm{Q}_1}^{x}-iD_{\bm{Q}_1}^{x} \\
E_{\bm{Q}_1}^{x}+iD_{\bm{Q}_1}^{x} & -E_{\bm{Q}_1}^{x}+iD_{\bm{Q}_1}^{x}  & F_{\bm{Q}_1}^{x}  
\end{pmatrix},\\
\label{eq:lambda_4}
X_{\bm{Q}_4}&=\begin{pmatrix}
F_{\bm{Q}_1}^{x} & -E_{\bm{Q}_1}^{x}-iD_{\bm{Q}_1}^{x} & -E_{\bm{Q}_1}^{x}+iD_{\bm{Q}_1}^{x} \\
-E_{\bm{Q}_1}^{x}+iD_{\bm{Q}_1}^{x} & F_{\bm{Q}_1}^{x}  & E_{\bm{Q}_1}^{x}+iD_{\bm{Q}_1}^{x} \\
-E_{\bm{Q}_1}^{x}-iD_{\bm{Q}_1}^{x} & E_{\bm{Q}_1}^{x}-iD_{\bm{Q}_1}^{x}  & F_{\bm{Q}_1}^{x}  
\end{pmatrix}.
\end{align}
$X_{\bm{Q}_2}$, $X_{\bm{Q}_3}$, and $X_{\bm{Q}_4}$ for $Pm\bar{3}m$, $P\bar{4}3m$, and $Pm\bar{3}$ are given by setting $D_{\bm{Q}_1}^{x}=0$ in Eqs.~(\ref{eq:lambda_2})--(\ref{eq:lambda_4}), respectively.

\subsection{$\{\bm{Q}\}_\Delta$}

The wave vectors $\bm{Q}_2$ and $\bm{Q}_3$ shown in Fig.~\ref{fig:BZ}(b) are connected to $\bm{Q}_1$ as $\bm{Q}_2=C^+_{3[111]}\bm{Q}_1$ and $\bm{Q}_3=C^-_{3[111]}\bm{Q}_1$.   
Then, $X_{\bm{Q}_2}$ and $X_{\bm{Q}_3}$ for $P23$ are given by
\begin{align}
\label{eq:delta_2}
X_{\bm{Q}_2}&=\begin{pmatrix}
F_{\bm{Q}_1}^{z} & 0 & -iD_{\bm{Q}_1}^{x}\\
0 & F_{\bm{Q}_1}^{x}  & 0 \\
iD_{\bm{Q}_1}^{x}& 0  & F_{\bm{Q}_1}^{y}
\end{pmatrix},\\
\label{eq:delta_3}
X_{\bm{Q}_3}&=\begin{pmatrix}
F_{\bm{Q}_1}^{y} & iD_{\bm{Q}_1}^{x} & 0 \\
-iD_{\bm{Q}_1}^{x} & F_{\bm{Q}_1}^{z}  & 0 \\
0& 0 & F_{\bm{Q}_1}^{x}
\end{pmatrix}.
\end{align}
$X_{\bm{Q}_2}$ and $X_{\bm{Q}_3}$ for $Pm\bar{3}m$ and $P\bar{4}3m$ are given by setting $F_{\bm{Q}_1}^{y}=F_{\bm{Q}_1}^{z}$ and $D_{\bm{Q}_1}^{x}=0$ in Eqs.~(\ref{eq:delta_2}) and (\ref{eq:delta_3}), respectively.
$X_{\bm{Q}_2}$ and $X_{\bm{Q}_3}$ for $P432$ are given by setting $F_{\bm{Q}_1}^{y}=F_{\bm{Q}_1}^{z}$ in Eqs.~(\ref{eq:delta_2}) and (\ref{eq:delta_3}), respectively.
$X_{\bm{Q}_2}$ and $X_{\bm{Q}_3}$ for $Pm\bar{3}$  are given by setting $D_{\bm{Q}_1}^{x}=0$ in Eqs.~(\ref{eq:delta_2}) and (\ref{eq:delta_3}), respectively.

\subsection{$\{\bm{Q}\}_\Sigma$}

The wave vectors $\bm{Q}_2$, $\bm{Q}_3$, $\bm{Q}_4$, $\bm{Q}_5$, and $\bm{Q}_6$ shown in Fig.~\ref{fig:BZ}(c) are connected to $\bm{Q}_1$ as $\bm{Q}_2=C^+_{3[111]}\bm{Q}_1$, $\bm{Q}_3=C^-_{3[111]}\bm{Q}_1$, $\bm{Q}_4=C_{2[100]}\bm{Q}_1$, $\bm{Q}_5=C^+_{3[111]}C_{2[100]}\bm{Q}_1$, and $\bm{Q}_6=C^-_{3[111]}C_{2[100]}\bm{Q}_1$. 
Then, $X_{\bm{Q}_2}$, $X_{\bm{Q}_3}$, $X_{\bm{Q}_4}$, $X_{\bm{Q}_5}$, and  $X_{\bm{Q}_6}$ for $P23$ are given by
\begin{align}
\label{eq:sigma_2}
X_{\bm{Q}_2}&=\begin{pmatrix}
F_{\bm{Q}_1}^{z} & iD_{\bm{Q}_1}^{y}  &-iD_{\bm{Q}_1}^{x}  \\
-iD_{\bm{Q}_1}^{y} & F_{\bm{Q}_1}^{x}  &  E_{\bm{Q}_1}^{z}  \\
 iD_{\bm{Q}_1}^{x}&  E_{\bm{Q}_1}^{z}   & F_{\bm{Q}_1}^{y}
\end{pmatrix},\\
\label{eq:sigma_3}
X_{\bm{Q}_3}&=\begin{pmatrix}
F_{\bm{Q}_1}^{y} & iD_{\bm{Q}_1}^{x}  &  E_{\bm{Q}_1}^{z}  \\
-iD_{\bm{Q}_1}^{x} & F_{\bm{Q}_1}^{z} & iD_{\bm{Q}_1}^{y}   \\
  E_{\bm{Q}_1}^{z} &   -iD_{\bm{Q}_1}^{y} & F_{\bm{Q}_1}^{x}
\end{pmatrix},\\
\label{eq:sigma_4}
X_{\bm{Q}_4}&=\begin{pmatrix}
 F_{\bm{Q}_1}^{x} & -E_{\bm{Q}_1}^{z} & iD_{\bm{Q}_1}^{y} \\
 -E_{\bm{Q}_1}^{z} & F_{\bm{Q}_1}^{y}  &  iD_{\bm{Q}_1}^{x} \\
 -iD_{\bm{Q}_1}^{y}& -iD_{\bm{Q}_1}^{x}  & F_{\bm{Q}_1}^{z}
 \end{pmatrix},\\
\label{eq:sigma_5}
X_{\bm{Q}_5}&=\begin{pmatrix}
 F_{\bm{Q}_1}^{z} & -iD_{\bm{Q}_1}^{y}  & - iD_{\bm{Q}_1}^{x} \\
  iD_{\bm{Q}_1}^{y} & F_{\bm{Q}_1}^{x}  &  -E_{\bm{Q}_1}^{z}  \\
  iD_{\bm{Q}_1}^{x} &    -E_{\bm{Q}_1}^{z} & F_{\bm{Q}_1}^{y}
\end{pmatrix},\\
\label{eq:sigma_6}
X_{\bm{Q}_6}&=\begin{pmatrix}
F_{\bm{Q}_1}^{y} & iD_{\bm{Q}_1}^{x}  & -E_{\bm{Q}_1}^{z}  \\
- iD_{\bm{Q}_1}^{x}& F_{\bm{Q}_1}^{z}  & -iD_{\bm{Q}_1}^{y}  \\
 -E_{\bm{Q}_1}^{z} & iD_{\bm{Q}_1}^{y}  & F_{\bm{Q}_1}^{x}
\end{pmatrix}.
\end{align}
$X_{\bm{Q}_2}$--$X_{\bm{Q}_6}$ for $Pm\bar{3}m$ are given by setting $F_{\bm{Q}_1}^{x}=F_{\bm{Q}_1}^{y}$ and $D_{\bm{Q}_1}^{x}=D_{\bm{Q}_1}^{y}=0$ in Eqs.~(\ref{eq:sigma_2})-(\ref{eq:sigma_6}), respectively.
$X_{\bm{Q}_2}$--$X_{\bm{Q}_6}$ for $P\bar{4}3m$ are given by setting $F_{\bm{Q}_1}^{x}=F_{\bm{Q}_1}^{y}$ and $D_{\bm{Q}_1}^{x}=-D_{\bm{Q}_1}^{y}$ in Eqs.~(\ref{eq:sigma_2})-(\ref{eq:sigma_6}), respectively.
$X_{\bm{Q}_2}$--$X_{\bm{Q}_6}$ for $P432$ are given by setting $F_{\bm{Q}_1}^{x}=F_{\bm{Q}_1}^{y}$ and $D_{\bm{Q}_1}^{x}=D_{\bm{Q}_1}^{y}$ in Eqs.~(\ref{eq:sigma_2})-(\ref{eq:sigma_6}), respectively.
$X_{\bm{Q}_2}$--$X_{\bm{Q}_6}$ for $Pm\bar{3}$ are given by setting $D_{\bm{Q}_1}^{x}=D_{\bm{Q}_1}^{y}=0$ in Eqs.~(\ref{eq:sigma_2})-(\ref{eq:sigma_6}), respectively.

\bibliography{main.bbl}
\end{document}